\documentclass[aps, prl,reprint,superscriptaddress,showpacs]{revtex4-1}

\usepackage[utf8]{inputenc}
\usepackage[dvips]{graphicx}
\usepackage{epsfig}
\usepackage{multirow}
\usepackage{arydshln}
\usepackage{nicefrac}
\usepackage{textcomp}
\usepackage{gensymb}
\usepackage{url}

\begin{document}

\title{Elastic softness of hybrid lead halide perovskites}

\author{A.C. Ferreira}
\affiliation{Laboratoire L{\'e}on Brillouin, CEA-CNRS, Universit\'e Paris-Saclay, CEA Saclay, 91191 Gif-sur-Yvette, France}
\affiliation{Laboratoire FOTON, INSA, Universit\'e Rennes, F–35708 Rennes France}
\author{A. L\'etoublon}
\affiliation{Laboratoire FOTON, INSA, Universit\'e Rennes, F–35708 Rennes France}
\author{S. Paofai}
\affiliation{Institut des Sciences Chimiques de Rennes, CNRS, Université de Rennes 1, Ecole Nationale Supérieure de Chimie de Rennes, INSA de Rennes, 35042 Rennes, France}
\author{S. Raymond}
\affiliation{Univ. Grenoble Alpes, CEA, INAC, MEM, 38000 Grenoble, France}
\author{C. Ecolivet}
\affiliation{Univ Rennes, CNRS, IPR [(Institut de Physique de Rennes)] UMR 6251, F-35000 Rennes, France}
\author{B. Ruffl\'e}
\affiliation{Laboratoire Charles Coulomb (L2C), UMR 5221 CNRS-Université de Montpellier, Montpellier, FR-34095, France}
\author{S. Cordier}
\affiliation{Institut des Sciences Chimiques de Rennes, CNRS, Université de Rennes 1, Ecole Nationale Supérieure de Chimie de Rennes, INSA de Rennes, 35042 Rennes, France}
\author{C. Katan}
\affiliation{Institut des Sciences Chimiques de Rennes, CNRS, Université de Rennes 1, Ecole Nationale Supérieure de Chimie de Rennes, INSA de Rennes, 35042 Rennes, France}
\author{M. I. Saidaminov}
\affiliation{King Abdullah University of Science and Technology (KAUST), KAUST Catalysis Center, KAUST Solar Center, Physical Sciences and Engineering Division (PSE), Thuwal 23955-6900, Saudi Arabia}
\author{A. A. Zhumekenov}
\affiliation{King Abdullah University of Science and Technology (KAUST), KAUST Catalysis Center, KAUST Solar Center, Physical Sciences and Engineering Division (PSE), Thuwal 23955-6900, Saudi Arabia}
\author{O. M. Bakr}
\affiliation{King Abdullah University of Science and Technology (KAUST), KAUST Catalysis Center, KAUST Solar Center, Physical Sciences and Engineering Division (PSE), Thuwal 23955-6900, Saudi Arabia}
\author{J. Even}
\affiliation{Laboratoire FOTON, INSA, Universit\'e Rennes, F–35708 Rennes France}
\author{P. Bourges}
\affiliation{Laboratoire L{\'e}on Brillouin, CEA-CNRS, Universit\'e Paris-Saclay, CEA Saclay, 91191 Gif-sur-Yvette, France}

\date{\today}

\begin{abstract}
 
Much recent attention has been devoted towards unravelling the microscopic optoelectronic properties of hybrid organic-inorganic perovskites (HOP).  Here we investigate by coherent inelastic neutron scattering spectroscopy and Brillouin light scattering, low frequency acoustic phonons in four different hybrid perovskite single crystals: MAPbBr$_3$, FAPbBr$_3$, MAPbI$_3$ and $\alpha$-FAPbI$_3$ (MA: methylammonium, FA: formamidinium). We report a complete set of elastic constants caracterized by a very soft shear modulus  C$_{44}$. Further, a tendency towards an incipient ferroelastic transition is observed in FAPbBr$_3$.  We observe a systematic lower sound group velocity in the technologically important iodide-based compounds compared to the bromide-based ones. The findings suggest that low thermal conductivity and hot phonon bottleneck phenomena are expected to be enhanced by low elastic stiffness, particularly in the case of the ultrasoft $\alpha$-FAPbI$_3$.

\end{abstract}

\maketitle

%---------------------------------------------------------------%
% However, the influence of their elastic properties on the charge-carrier dynamics is currently still lacking a comprehensive understanding.

Motivated by environmental and energy issues, hybrid organolead perovskites (HOP) have drawn a lot of interest in the field of photovoltaic cells \cite{kojima2009organometal, mitzi2001structurally, knutson2005tuning, park2015perovskite, yang2015high, anaraki2016highly, NREL}. Currently, the state-of-the-art of HOP solar cells is based on alloys where methylammonium (MA) and formamidinium (FA) are both present in the same structure and \textit{ca.} 10\% are replaced by rubidium (Rb) and caesium (Cs) atoms, together with concomitant alloying of iodide (I)/bromide (Br) halogens \cite{saliba2016incorporation}, now reaching power conversion efficiencies (PCEs) in excess of 22\% \cite{yang2017iodide}. The rapid emergence and success of hybrid perovskites is widely attributed to various features such as low cost and low temperature processing, suitable optical bandgap (especially for iodide based HOPs), superb optical absorption across the visible spectrum, low exciton binding energies and long charge-carrier diffusion lengths.
Much recent attention has been devoted towards HOPs charge-carrier features \cite{yamada2014photocarrier, d2014excitons, wehrenfennig2014high, herz2016charge} and previous experimental studies have successfully evidenced the influence of crystal structure on their optoelectronic properties, as exemplified with the effect of octahedral tilt on the band-gap \cite{even2013importance, even2014analysis}. However, the origin of the softness of HOPs, compared to classic semiconductors, and its influence on their charge-carrier dynamics is still lacking a comprehensive understanding and systematic experimental studies. Furthermore, in perovskite based thin film solar cells, device performance is deeply affected by film quality and fabrication processing (morphological effects, grain boundaries, etc.), which makes it difficult to study the intrinsic properties of HOPs. Single crystals, on the other hand, provide the ideal platform to uncover their fundamental limits.
In this work, we have investigated low frequency structural excitations in the cubic phases of the most relevant compounds implemented in HOPs, namely MAPbBr$_3$, FAPbBr$_3$, MAPbI$_3$ and $\alpha$-FAPbI$_3$, in their single crystal form. For additional information regarding structure and crystal growth please refer to the Supplemental Material \cite{SuppMat}. We report a complete set of elastic constants, via the corresponding sound velocities, and we relate the results to the lower thermal conductivity found in HOP compounds and the hot phonon bottleneck hypothesis proposed for such systems. For that purpose, dispersions of the acoustic phonons have been measured around main Bragg reflections using inelastic neutron scattering (INS). Complementary Brillouin light scattering (BLS) experiments have also been used to determine sound velocities in the bromide-based compounds. The experimental conditions and procedure of both techniques are detailed in the Supplemental Material \cite{SuppMat}.

%---------------------------------------------------------------%

%\section*{Results}

% \subsection*{Low energy INS spectra}

\begin{figure}
\centering
\includegraphics[height=7.75cm, keepaspectratio=true]{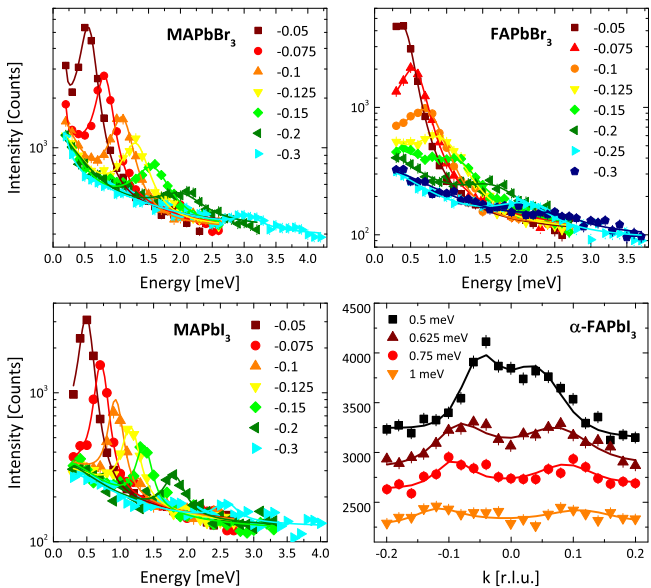}
\caption{
Transverse acoustic (TA) phonon spectra measured by inelastic neutron scattering in the cubic phase of (a) MAPbBr$_3$), (b) FAPbBr$_3$, (c) MAPbI$_3$ and (d) $\alpha$-FAPbI$_3$, for different Q positions going away from the (002)$\equiv$(200) Bragg peak (\textit{i.e.} Q = (k k 2)) and for different energy values. With the exception of MAPbI$_3$ (340 K), all other compounds were studied at room temperature (RT).
}
\label{fig:INS-Spectra-TA002}
\end{figure}

\textbf{Fig. \ref{fig:INS-Spectra-TA002}} shows the low energy INS spectra of transverse acoustic (TA) phonons in MAPbBr$_3$, FAPbBr$_3$, MAPbI$_3$ and $\alpha$-FAPbI$_3$, measured around the Bragg reflection (200). Additional data around other Bragg reflections, (110) and (111), can be found in the Supplemental Material \cite{SuppMat}. We performed both constant energy and Q scans at the main Bragg positions. Using the (200) position as an example, longitudinal (LA) and transverse (TA) acoustic modes were measured at different reciprocal space positions Q = (200) + $q$ in HKL units, with $q$ along (LA) and perpendicular (TA) to [200] (for TA $q$ is parallel to [011]). Clear acoustic phonon modes are observed on top of a background. As shown by constant Q-scans, the background results from large quasi-elastic signals, likely coming from incoherent scattering of hydrogen atom excitations present in the MA/FA molecules, as it has been previously studied on powder samples in MAPbBr$_3$ \cite{Swaissonneutrons}. Energy scans performed at various momentum vectors around the Bragg positions revealed that the quasi-elastic signal occurring in the energy window of the acoustic phonons is almost independent of the sample orientation and weakly dependent of the longitudinal momentum. The quasi-elastic signal corresponds to a flat contribution in constant-energy scans, whereas acoustic phonons show up as a double peak structure at symmetric positions of the Bragg reflection (\textit{e.g.} \textbf{Fig. \ref{fig:INS-Spectra-TA002}(d)}). As a result, even though the background is relatively large, one can easily separate it from the dispersing and symmetric phonon modes. All neutron spectra have been fitted as described in the Supplemental Material \cite{SuppMat}, from which the phonon energy can be extracted. 

\begin{table*}
\centering
\scriptsize
\def\arraystretch{1.5}
\begin{ruledtabular}
\caption{Summary of the elastic properties at RT for MAPbBr$_3$, FAPbBr$_3$,  and $\alpha$-FAPbI$_3$  and at 340 K for MAPbI$_3$, as measured by Inelastic Neutron Scattering (INS) and Brillouin Scattering (BS). In  $\alpha$-FAPbI$_3$, the bulk modulus is found: K= 0.1 $\pm$ 2.3 GPa (we here only quote the positive range which is only  physically meaningfull).}
  \begin{tabular}{ccccccccc}
\multirow{2}{*}{\textbf{Elastic Constant}} & \multicolumn{2}{c}{\textbf{MAPbBr$_3$}} & \multicolumn{2}{c}{\textbf{FAPbBr$_3$}} & \multicolumn{2}{c}{\textbf{MAPbI$_3$}} & \multicolumn{2}{c}{\textbf{FAPbI$_3$}}     \\
                                  & INS         & BS         & INS         & BS         & INS         & BS         & INS & \multicolumn{1}{l}{BS} \\ \hline
 \boldmath{$\mathrm{C_{11}}$}{[GPa]}     &     34.5$\pm$1.2      &    32.2$\pm$0.2 & 27.7$\pm$1.6 & 31.2$\pm$0.2 & 21.8$\pm$1.3 & n/a & 11.1$\pm$2.0 & n/a    \\
\boldmath{$\mathrm{C_{44}}$}{[GPa]}     &     4.1$\pm$0.2 & 3.4$\pm$0.1 & 3.1$\pm$0.1 & 1.5$\pm$0.1 & 7.3$\pm$0.3 & n/a & 2.7$\pm$0.3 & n/a  \\
\boldmath{$\mathrm{C_{12}}$}{[GPa]}   &   18.5$\pm$2.0 & 9.1$\pm$0.8 & 11.5$\pm$2.4 & 9.4$\pm$0.5 & 11.3$\pm$3.1 & n/a & -5.5$\pm$2.2 & n/a  \\ \hline
\textbf{Bulk modulus} {[GPa]}   & 23.9$\pm$1.3 &   16.8$\pm$0.1   & 16.9$\pm$1.7  &   16.667$\pm$0.3  &  14.8$\pm$1.7  &   n/a  & 0.0-2.4  &    n/a     \\
\textbf{Anisotropy, A}    &   0.52$\pm$0.0.05  & 0.29$\pm$0.01  & 0.38$\pm$0.03 & 0.14$\pm$0.01  &   1.38$\pm$0.22  &  n/a  & 0.4$\pm$0.2 &  n/a         \\
\textbf{L/T ratio}       & 8.7$\pm$0.5  & 9.5$\pm$0.3  & 8.9$\pm$0.7  &  20.8$\pm$1.4  & 3.0$\pm$0.2  &   n/a   & 4.3$\pm$0.9  &  n/a  
 \end{tabular}
  \end{ruledtabular}
\label{tab:elastic-constants}
\end{table*}

\begin{figure*}
\centering
\includegraphics[width=17cm, keepaspectratio=true]{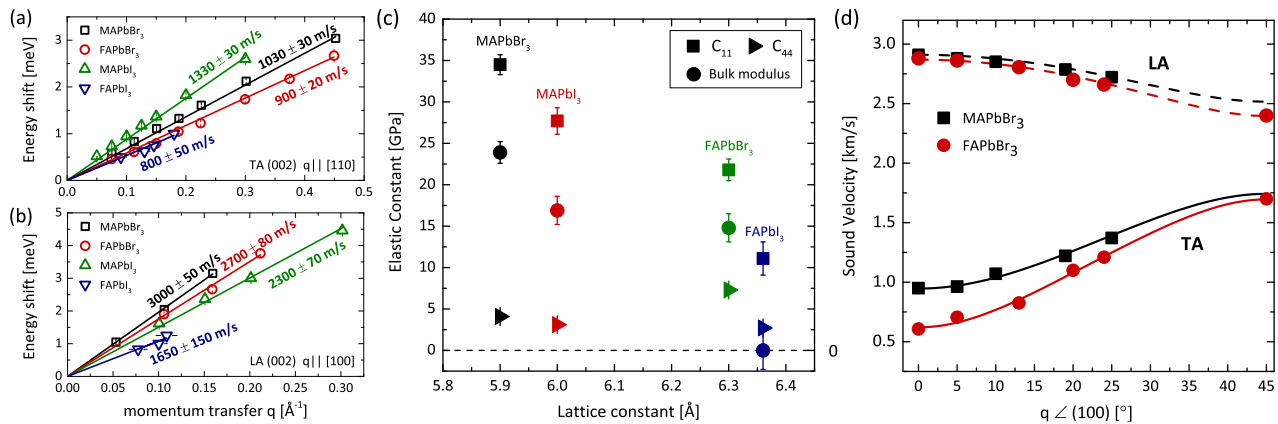}
\caption{
Acoustic phonon dispersion curves of the four HOP (a) TA and (b) LA phonons close to the (002) Bragg reflection, measured by INS. The data for a given configuration ($q$ direction) is regrouped for the different HOP systems and differentiated by different colors and symbols. (c) Elastic constants C$_{11}$ and C$_{44}$ (black) as well as the bulk modulus K (red) behaviour as a function of the changing lattice constant between compounds. (d) Sound velocity diagram for FAPbBr$_3$ and MAPbBr$_3$, as determined by Brillouin light scattering. The velocity is given as a function of the angle between the direction of measurement and the [100] direction. }
\label{fig:Dispersions-C11-C44-Bulk-BS}
\end{figure*}

By varying the distance $q$ to the nearest Bragg peak, \boldmath{$\mathit{\Gamma}$} point, one can draw the dispersion curves, which are reported in \textbf{Fig. \ref{fig:Dispersions-C11-C44-Bulk-BS}} and in Supplemental Material \cite{SuppMat}. Each sub-figure represents measurements at a different Bragg position or propagation direction. We derive the phonon sound velocity V from linear regressions along most directions and then, with simple rules of elasticity in cubic systems \cite{halliday1997fundamentals}, readily determine the corresponding elastic constants for the four perovskite compounds. \textbf{Fig. \ref{fig:Dispersions-C11-C44-Bulk-BS}(a) and  \ref{fig:Dispersions-C11-C44-Bulk-BS}(b)} specifically show the sound velocities of transverse and longitudinal acoustic phonons around the (002) Bragg reflection. These respectively yield C$_{44}$ and C$_{11}$ in a rather direct way ($C=\rho$V$^2$) which are then plotted as a function of the lattice constant in \textbf{Fig. \ref{fig:Dispersions-C11-C44-Bulk-BS}(c)}. A summary of the determined sound velocities by INS for all four HOP compounds is given in the Supplemental Material \citep{SuppMat} and the extracted elastic constants are in \textbf{Table 1\ref{tab:elastic-constants}}. Considering the obtained elastic constants we can also calculate the bulk modulus \textbf{K} $\rm=\nicefrac{1}{3}(C_{11}+2C_{12})$ \cite{halliday1997fundamentals}, Zener anisotropy index \textbf{A} $\rm=\nicefrac{2C_{44}}{(C_{11}-C_{12})}$ \cite{zener1948elasticity} and longitudinal/transverse (L/T) ratio for all systems (see Supplemental Material \citep{SuppMat}).

Immediately we note that C$_{12}$ in $\alpha$-FAPbI$_3$ is negative, but still it respects the necessary Born elastic stability criteria for cubic systems  \cite{born1940stability, mouhat2014necessary}. A negative C$_{12}$ simply implies that a cubic material, when uniaxially compressed along a [100] direction, will contract in the other two directions ([010] and [001]) and in that way, try to maintain an isotropic structure.  Nonetheless, the negative C$_{12}$ together with the very low bulk modulus, confirms the very unstable nature of $\alpha$-FAPbI$_3$ (whose metastable single crystals last less than a week in the $\alpha$ phase \cite{zhumekenov2016formamidinium}) and why it has actually been paired with MA, Rb and Cs for better performing photovoltaic devices. As shown in \textbf{Table 1\ref{tab:elastic-constants}}, the four perovskite compounds exhibit an overall sizeable elastic anisotropic nature (A$\neq$1), which can be mostly attributed to the very low shear modulus C$_{44}$. MAPbI$_3$ stands out with its relatively higher (around double) C$_{44}$ which results in a discrepant, although still anisotropic, Zener index. However, it should be noted that the measurements were performed at 340 K, just above the transition to the cubic phase \cite{whitfield2016structures}, resulting in an additional anharmonicity effect. The remarkably low shear moduli are much more evident when compared to the ones of classical semiconducting materials such as GaAs, AlAs, etc., where the elastic constants are in the 10$^2$ GPa order \cite{vurgaftman2001band}. It is believed that the rotation/tilts of the corner sharing PbX$_6$ octahedra \cite{swainson2007pressure} could be responsible for the particularly low shear modulus and large anisotropy in HOPs. 

As evidenced in \textbf{Fig. \ref{fig:Dispersions-C11-C44-Bulk-BS}(c)}, C$_{11}$ and K decrease noticeably with increasing lattice constant. These two quantities are, therefore, lower in iodide-based systems compared with bromide ones, especially in $\alpha$-FAPbI$_3$ where they are a third of its bromide counterpart. This indicates a structural instability when the lattice parameter exceeds 6.4 \AA. The relatively higher bulk modulus in MA-based compounds should be related to a steric effect, where the more symmetric and rotating MA molecules lead to more compact structures, which in turn results in larger binding elastic interactions. It is worth emphasizing that, by construction, the acoustic branches are defining the lowest zone boundary phonons and thus, related to the lowest peaks of the phonon density of states. That also implies that the Debye temperature is very small (\textit{i.e.} about 30 K). More, as a function of lattice constant, this quantity exhibits the same trend as for C$_{11}$ and the bulk modulus. The observation that iodide materials are softer than bromide materials, as well as FA-based compounds versus MA-based ones is consistent with recent static nanoindentation measurements of the Young modulus \cite{rakita2015mechanical, elbaz2017phonon}, though $\alpha$-FAPbI$_3$ was not explored (see table in Supplemental Material \cite{SuppMat} for comparison). In contrast, recent pulse-echo ultrasonic measurements at low frequency (10 MHz) and in the 140-350 K temperature range \cite{anusca2017dielectric}, show larger sound velocities for MAPbI$_3$ than for MAPbBr$_3$. However, the results of Anusca \textit{et al.} \cite{anusca2017dielectric} are also in disagreement with other ultrasonic studies \cite{letoublon2016elastic, lomonosov2016exceptional}. Yet it should be noted that the sound velocity attenuation is very large at 10 MHz and strongly affected by the structural phase transitions. Brillouin light scattering on the other hand, allows exploring the same properties in the GHz range, intermediate between ultrasonic and neutron scattering measurements. 

\begin{figure}[t]
\centering
\includegraphics[width=7cm, keepaspectratio=true]{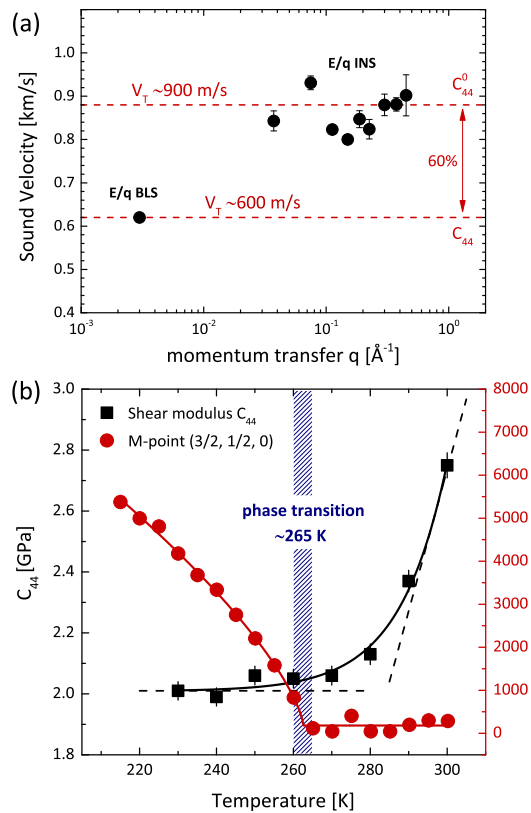}
\caption{
Softening of C$_{44}$ in FAPbBr$_3$ (a) as function of $q$ (at RT) and (b) as a function of temperature (230 - 300 K), around the (002) Bragg reflection. Also in (b) the Bragg M point intensity (red) across a similar a temperature range. The bare elastic constant, C$^0_{44}$, represents the elastic properties without the influence of the (incipient) phase transition. Note that the $q$ scale is logarithmic to underline the broad q-range covered by both experimental techniques.}
\label{fig:FAPB-Softening}
\end{figure}

Sound velocities were measured in both bromides compounds at room temperature (RT) using Brillouin light scattering, with a set-up in the [100], [010] base plane \cite{letoublon2016elastic}. Five different incidence angles between the normal (0\degree) and the Brewster angle (25\degree) are reported in \textbf{Fig. \ref{fig:FAPB-Softening}(a)} with the observation of quasi longitudinal and quasi transverse acoustic modes. A measurement along the cubic diagonal [110] is also shown. For both bromide compounds, a good agreement between INS and BLS longitudinal sound velocities is observed (see Supplemental Material \cite{SuppMat}). The same applies for the transverse mode in MAPbBr$_3$ \cite{letoublon2016elastic},  within  a 5\% difference. In contrast, a 30\% difference is observed for the same mode in FAPbBr$_3$. This is emphasized in \textbf{Fig. \ref {fig:FAPB-Softening}(b)} where the sound velocity is presented as a function of $q$. One can clearly observe the phonon softening of the transverse mode at lower $q$, in the BLS regime, which indicates a tendency towards a ferroelastic phase transition \cite{cummins}. Recording that this specific sound velocity is related to the C$_{44}$ elastic constant as C$_{44}$=$\rho$V$^2$ (where $\rho$= 4087 kg/cm$^3$ is the density of FAPbBr$_3$), it means that a 30\% softening of the sound velocity corresponds to a 60\% re-normalization of C$_{44}$. Interestingly, the values obtained by laser ultrasonics \cite{lomonosov2016exceptional}, where a set of elastic constants was also given for MAPbBr$_3$, are in good agreement with our BLS results and a similar softening of C$_{44}$ is observed too. 

The softening of the C$_{44}$ elastic constant is typically related to the proximity of a ferroelastic transition  \cite{cummins} which, in this case, can be only incipient. To test that possibility, we have performed a temperature study of the acoustic branch using INS. A softening of acoustic phonon at \textbf{Q} $=(2,0.025,0.025)$ is indeed observed right below RT in FAPbBr$_3$. As shown in \textbf{Fig. \ref{fig:FAPB-Softening}(b)}, the shear modulus C$_{44}$ is drastically reduced upon cooling, however, that effect is stabilized below  $\sim$270 K. Actually, this temperature range corresponds to the growing of an additional scattering in elastic neutron scattering at the M point, like \textit{e.g.} \textbf{Q} $=(\nicefrac{3}{2},\nicefrac{1}{2},0)$ (\textbf{Fig. \ref{fig:FAPB-Softening}(b)}). The measurements of the M point intensity show the appearence of a phase transition at 263 K in FAPbBr$_3$ towards a tetragonal structure, caracterized by a doubling of the cubic unit cell. In general, lead perovskites are indeed known to exhibit structural instabilities at both the M and R points \cite{fujii1974neutron}. From Fig. \ref{fig:FAPB-Softening}.b, one can see that the phase transition is blocking the development of the ferroelastic instability, limiting pre-transitional effects at high temperature, compared to what would be expected for a true ferroelastic transition \cite{cummins}. By linearly extrapolating the high temperature behaviour, we can estimate that such a ferroelastic transition would occur at 240 $\pm$ 20 K, barring the occurrence of the structural instability at the M point. Consistently with an aborted ferroelastic instability, a very modest phonon broadening is observed with decreasing temperature from 300 to 230 K. %  (see \textbf{Fig. \ref{fig:FAPB-Softening}(c)}).

%---------------------------------------------------------------%

%\section*{Discussion}

HOPs are characterized by high electron/hole free charge carrier mobility at room temperature. At high temperature ranges such as RT (10 times the Debye temperature), the electronic mobility, $\mu$, is typically governed by phonon scattering, via electron-phonon coupling (Fröhlich phonon emission). The whole phonon spectrum will then contribute to the electronic scattering rate, 1/$\tau$ ($\mu$ is proportional to the electronic relaxation time $\tau$). However, for intra-valley electron bands of direct gap semiconductors like the 3D HOP, the scattering rate should also be enhanced by collisions with low energy longitudinal acoustic phonons. When the acoustic phonon contribution is considered, the electronic scattering time, $\tau$, is expected to be proportionnal to the average squared longitudinal sound velocities \cite{Chambers}, \textit{i.e.} proportional to the average elastic constants such as $C_{11}$ or the bulk modulus $K$ that are shown in \textbf{Fig. \ref{fig:Dispersions-C11-C44-Bulk-BS}(c)}. The contributions to the carrier mobilities related to interactions with acoustic phonons are thus predicted to be strongly different between iodide- and bromide-based compounds, but such a large difference is not experimentally observed here. Instead, it further shows that the carrier mobilities are rather limited by other processes, namely interactions with optical phonons. Such observation is corroborated by emission line broadening results \cite{wright2016electron, diab2016narrow}, and it confirms that scattering from longitudinal optical phonons is the dominant source of electron-phonon coupling near RT.  

It is necessary, however, to consider the acoustic phonons in order to explain other optoelectronic properties that Fröhlich optical phonon emission does not account for. More precisely, a significant hot-phonon bottleneck effect in carrier thermalization is observed in lead-halide perovskites \cite{yang2016observation, yang2017acoustic}, when carrier injection levels are high. This is explained by the up-conversion of acoustic phonons which recycle thermal (vibrational) energy, reheating charge carriers and prolonging the cooling period of carrier-optical phonon system. Ultrafast transient absorption measurements reveal two stages of the carrier cooling process \cite{yang2017acoustic}. The first one is related to the intrinsic Fröhlich phonon emission mentioned above and does not vary significanlty among the various different peroskites. On the other hand, in the second cooling stage, hot carrier-phonon dynamics occur, corresponding to a phonon bottleneck effect \cite{yang2016observation, yang2017acoustic}. The various HOP materials studied in the present work show, in fact, very different accoustic phonon densities of states that might affect hot phonon energy relaxations. At RT, the carrier-phonon relaxation rate of that second cooling stage is typically 3 times slower in $\alpha$-FAPbI$_3$, compared with the MAPbBr$_3$ system \cite{yang2017acoustic}. This is fully consistent with the difference in elastic constants reported in \textbf{Fig. \ref{fig:Dispersions-C11-C44-Bulk-BS}(c)}.

Another direct consequence of the lattice softness is the ultralow thermal conductivities reported in perovskites \cite{wang2016}. Thermal conductivity is typically proportional to the square of the phonon sound group velocity \cite{Chambers, wang2016}, corresponding to an average elastic constant (\textbf{Fig. \ref{fig:Dispersions-C11-C44-Bulk-BS}(c)}). One can then associate lower thermal conductivity to lower elastic stiffness \cite{pisoni2014ultra, kovalsky2017thermal, elbaz2017phonon}. It is worth mentioning that a strong phonon bottleneck effect is essential for hot carrier photovoltaic devices, as it allows for a long-lived hot carrier population. Lower elastic bulk modulus, sign of low thermal conductivity, appears to be, therefore, important to enhance the hot carrier effect.

In conclusion, our quantitative study of low energy acoustic phonons presents a complete set of elastic constants of various technologically relevant hybrid perovskites, in their cubic phases, and shows a noticeable variation of elastic bulk modulus among them. We give a clear explanation for the ultralow thermal conductivities measured in hybrid organolead perovskites. Moreover, the data presented here strongly supports the hypothesis of the hot phonon bottleneck phenomena, reported by other authors to explain hot carriers relaxations. Both processes are expected to be enhanced by low elastic stiffness, especially in the case of the ultrasoft $\alpha$-FAPbI$_3$.

%---------------------------------------------------------------%

\bibliography{Biblioperov}

\section*{Supplemental Material}
 
\cleardoublepage

\clearpage
\includegraphics[width=17 cm]{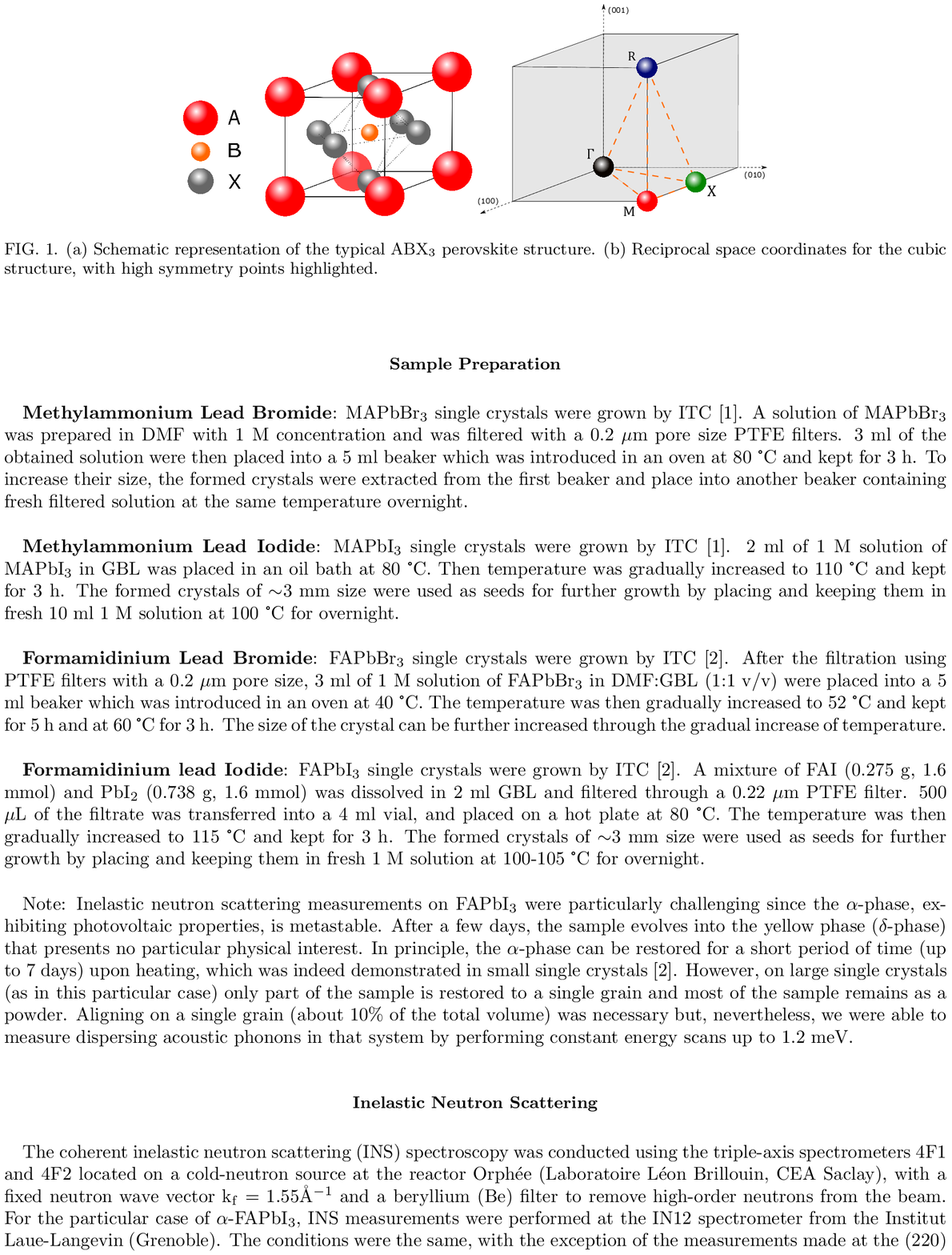}
\clearpage
\includegraphics[width=17cm]{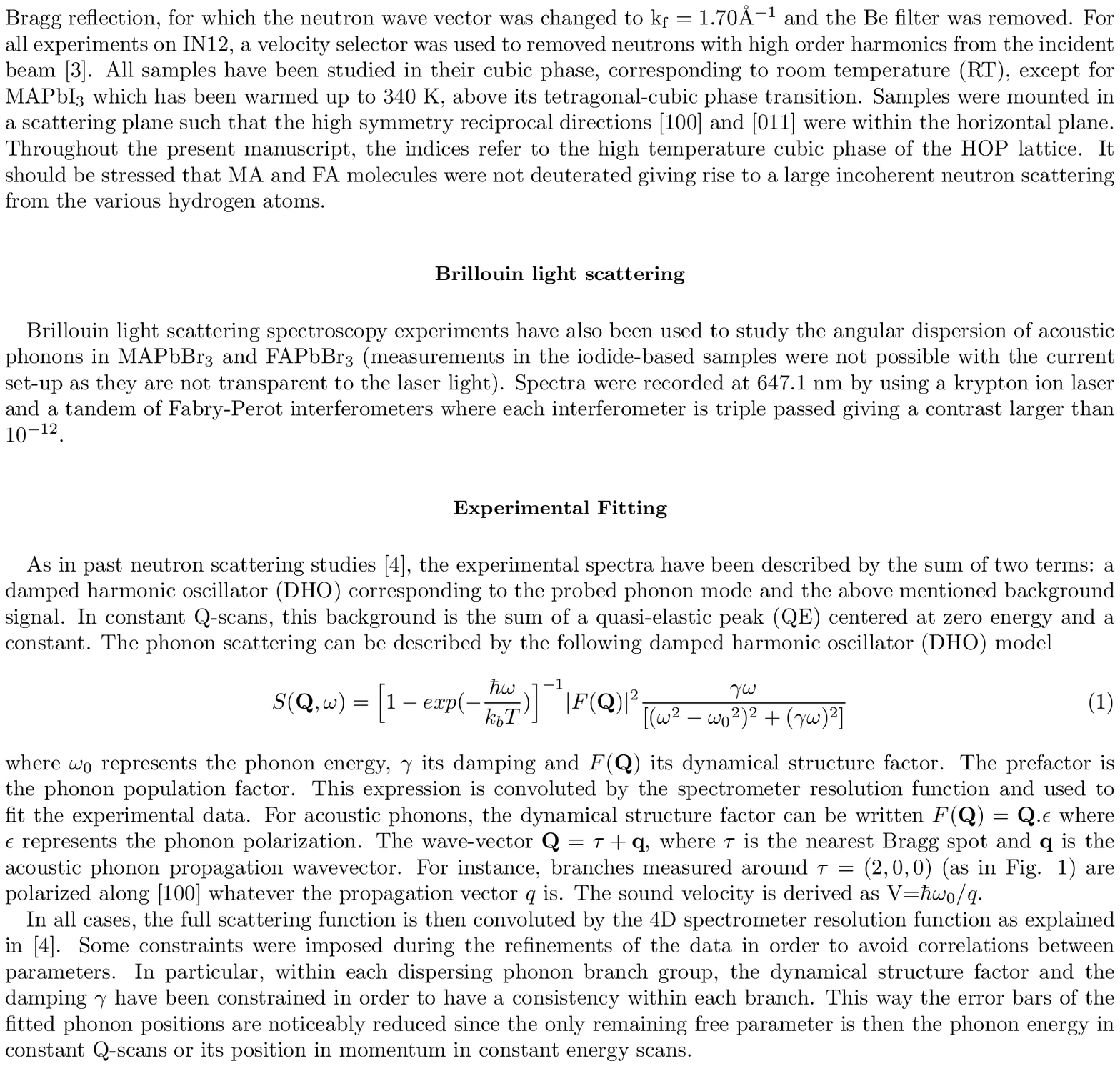}
\clearpage
\includegraphics[width=17 cm]{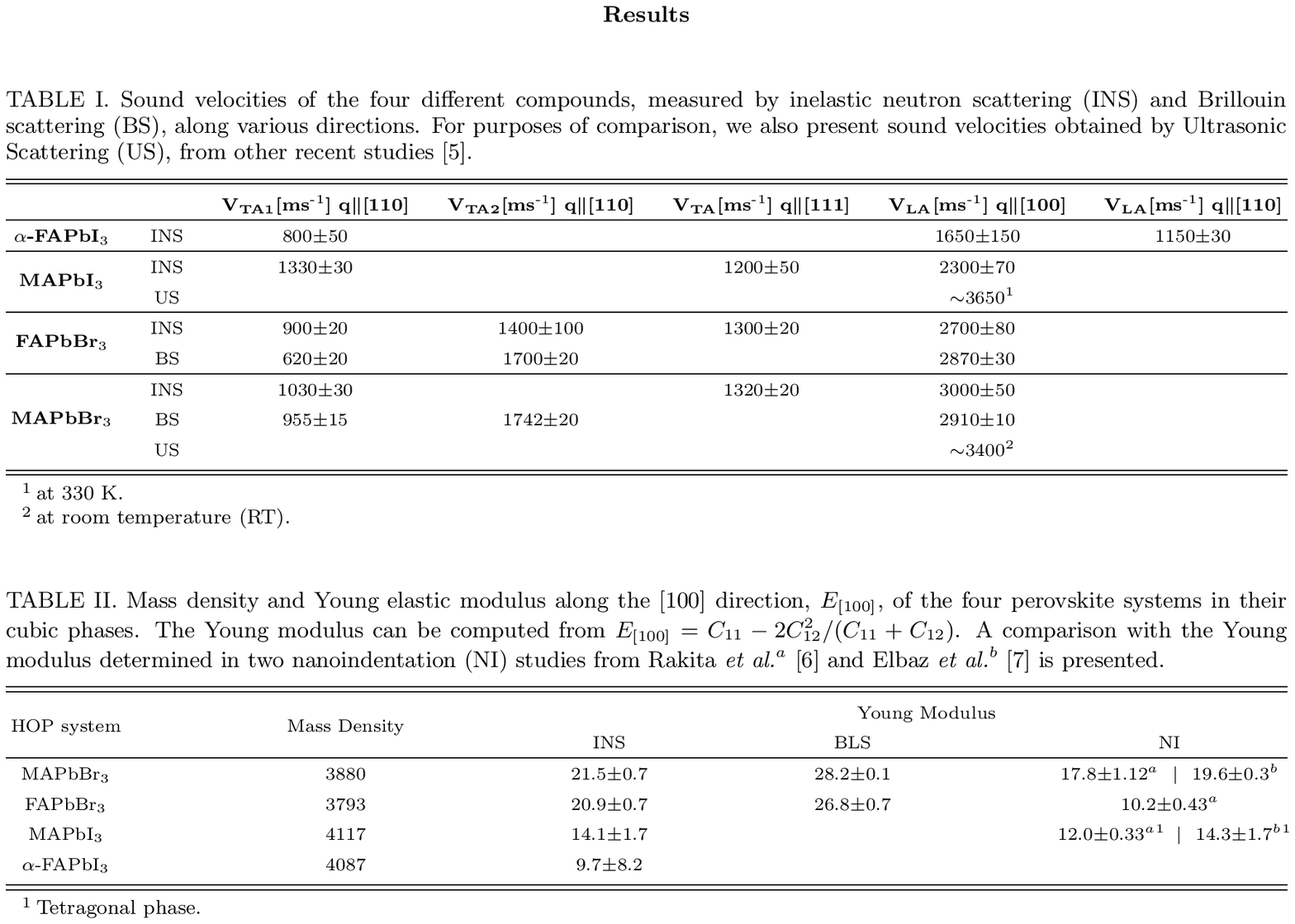}
\clearpage
\includegraphics[width=17 cm]{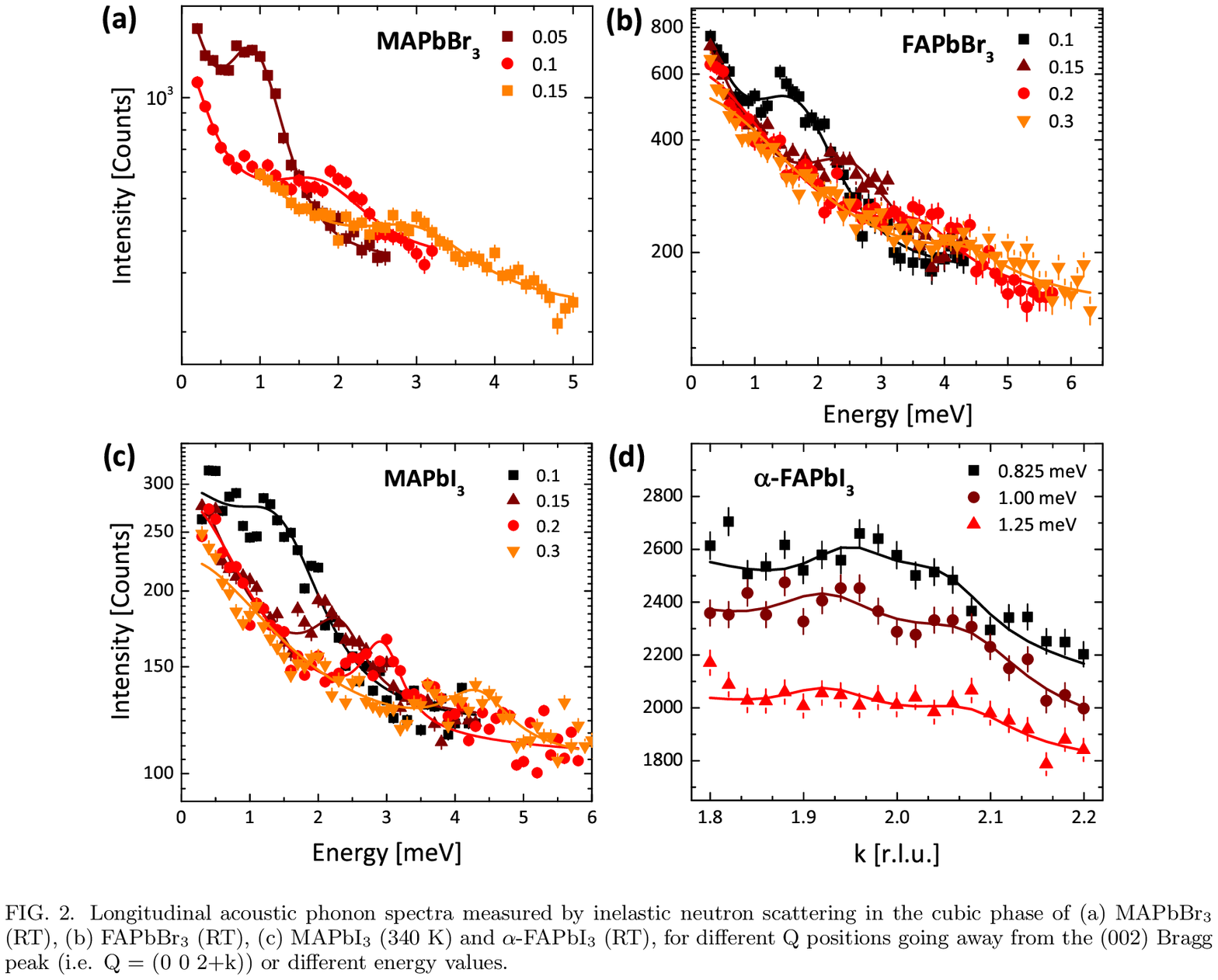}
\clearpage
\includegraphics[width=17cm]{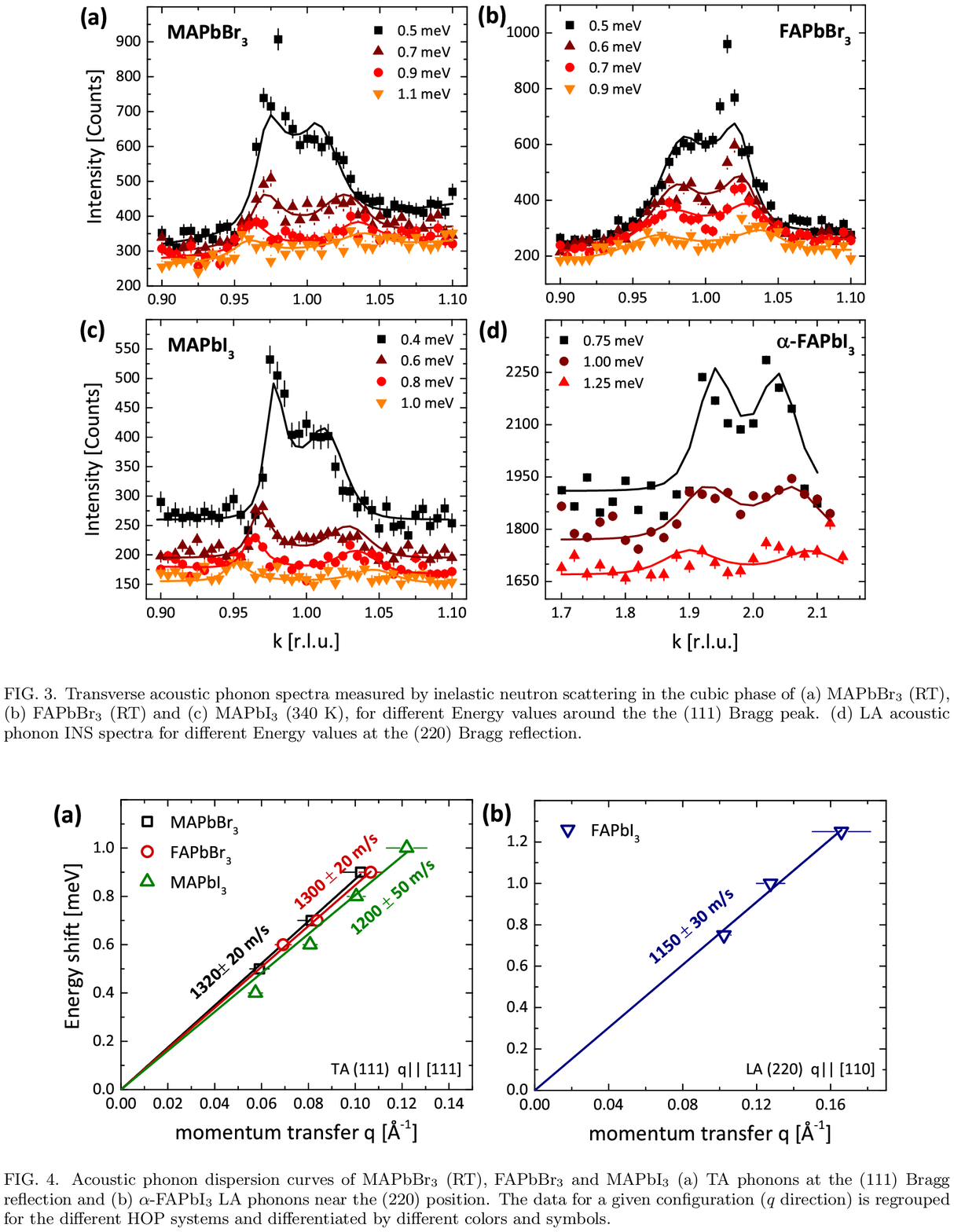}
\clearpage
\includegraphics[width=17 cm]{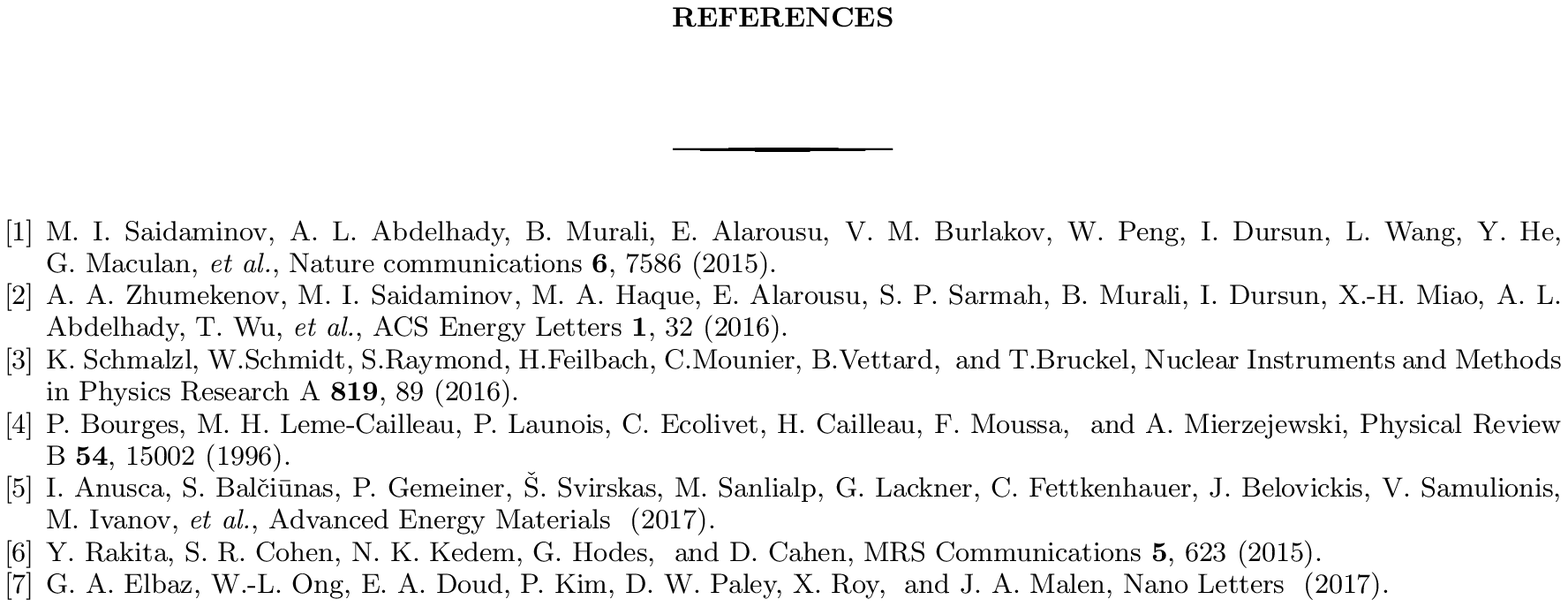}

-----------------------------------------
\end{document}